\documentclass[aps,amssymb,floatfix,superscriptaddress,preprintnumbers,nofootinbib]{revtex4}

\usepackage{graphicx}
\usepackage{bm}
\usepackage{amsmath}
\usepackage{amssymb}
\usepackage{amsfonts}
\usepackage{float}
\usepackage{hyperref}
\usepackage{dsfont}  
\usepackage{slashed}  
\usepackage{booktabs}
\usepackage{multirow}
\usepackage{subfigure}
\usepackage[sort&compress]{natbib}
\usepackage{xcolor}
\usepackage{ulem}

\newcommand{\be}{\begin{equation}}  
\newcommand{\ee}{\end{equation}}

\newcommand{\bea}{\begin{eqnarray}}
\newcommand{\eea}{\end{eqnarray}}

\newcommand{\Slash}[1]{{\ooalign{\hfil/\hfil\crcr$#1$}}}

\newcommand{\nn}{\nonumber\\}
\newcommand{\beq}{\begin{eqnarray}}
\newcommand{\eeq}{\end{eqnarray}}

\begin{document}

\title{Chiral and trace anomalies in Deeply Virtual Compton Scattering}

\author{Shohini Bhattacharya}
\email{sbhattach@bnl.gov}
\affiliation{Physics Department, Brookhaven National Laboratory, Upton, NY 11973, USA}

\author{Yoshitaka Hatta}
\email{yhatta@bnl.gov}
\affiliation{Physics Department, Brookhaven National Laboratory, Upton, NY 11973, USA}
\affiliation{RIKEN BNL Research Center, Brookhaven National Laboratory, Upton, NY 11973, USA}

\author{Werner Vogelsang}
\email{werner.vogelsang@uni-tuebingen.de}
\affiliation{Institute for Theoretical Physics, T\"{u}bingen University, Auf der Morgenstelle 14, 72076 T\"{u}bingen, Germany}

\begin{abstract}

\vspace*{0.4cm}
\noindent Inspired by recent work by Tarasov and Venugopalan, we calculate the one-loop quark box diagrams relevant to polarized and unpolarized Deep Inelastic Scattering (DIS) by introducing off-forward momentum $l^\mu$ as an infrared regulator. 
In the polarized case, we rederive the pole  $1/l^2$ related to the axial (chiral) anomaly. In addition, we obtain the usual logarithmic term and the DIS coefficient function. We interpret the result in terms of the generalized parton distributions (GPDs) $\tilde{H}$ and $\tilde{E}$ and discuss the possible violation of QCD factorization for the Compton scattering amplitude. Remarkably, we also find poles in the unpolarized case which are remnants of the  trace anomaly. We argue that these poles are canceled by the would-be massless  glueball poles in the GPDs $H$ and $E$ as well as in  their moments, the nucleon gravitational form factors $A,B$ and $D$. This mechanism sheds light on the connection between the gravitational form factors and the gluon condensate operator $F^{\mu\nu}F_{\mu\nu}$.  

\end{abstract}

\maketitle

\section{Introduction}
The role of the $U_A(1)$ axial (chiral) anomaly in polarized Deep Inelastic Scattering (DIS) has a long and winding history. Originally in the late 1970s it was used to constrain the one-loop gluonic  correction to the first moment of the singlet $g_1(x)$ structure function, as well as the two-loop anomalous dimension of the quark helicity contribution $\Delta \Sigma$ to the proton spin~\cite{Kodaira:1979pa}. Soon after the discovery of the `spin crisis' by the European Muon Collaboration (EMC) in 1988~\cite{EuropeanMuon:1987isl}, it was suggested that an anomaly-induced gluon helicity $\Delta G$  contribution to the `intrinsic' quark helicity $\Delta\tilde{\Sigma}$ 
\beq
\Delta \Sigma = \Delta \tilde{\Sigma}-\frac{n_f\alpha_s}{2\pi}\Delta G \, ,
\label{crisis}
\eeq
could be the key to explaining the unexpectedly small value of $\Delta \Sigma$~\cite{Carlitz:1988ab, Altarelli:1988nr}. While  such a scenario became popular at the time, subsequent decades-long experiments and global analyses did not find evidence for a sufficiently large $\Delta G$ to make it phenomenologically viable~\cite{deFlorian:2014yva,Nocera:2014gqa,Zhou:2022wzm}.
More importantly, the identification of the axial anomaly contribution as gluon helicity also met with much theoretical objection from the very start~\cite{Jaffe:1989jz,Forte:1989qq,Bodwin:1989nz,Vogelsang:1990ug}. The gluonic contribution in~(\ref{crisis}) comes from the infrared region of the triangle diagram which contains the Adler-Bell-Jackiw anomaly.
As pointed out by Jaffe and Manohar~\cite{Jaffe:1989jz}, a proper way to regulate the infrared singularity of this diagram is to calculate it in {\it off}-forward kinematics. The anomaly then manifests itself as a pole in momentum transfer $l=p_2-p_1$ in the matrix element of the singlet axial current $J_5^\mu= \sum_f\bar{\psi}_f \gamma^\mu \gamma_5\psi_f$,
\beq
\langle p_2|J_5^\mu|p_1\rangle = \frac{n_f\alpha_s}{4\pi}\frac{il^\mu}{l^2}\langle p_2|F_a^{\alpha\beta}\tilde{F}_{\alpha\beta}^a|p_1\rangle \, ,
\label{trian}
\eeq
where the incoming and outgoing gluons are on shell and the $n_f$ quarks in the loop are massless. The appearance of the pole $1/l^2$ and the twist-four pseudoscalar operator $F\tilde{F}$ seem alarming, as they signal some underlying nonperturbative physics that does not fit into the standard perturbative QCD framework. Yet, in ordinary perturbative calculations done in forward kinematics, this problem is superficially avoided  by certain  choices of infrared regularization such as dimensional regularization. 

Decades after the initial controversy, infrared sensitivity and the subtleties of taking the forward limit seemed to have been largely forgotten. Nowadays, forward kinematics is  routinely used in the  higher-order computations of polarized cross sections and asymmetries. However, recently the issue of the anomaly pole has been  rekindled by Tarasov and Venugopalan~\cite{Tarasov:2021yll,Tarasov:2020cwl} who pursued and crystallized the original suggestion by Jaffe and Manohar. They have demonstrated, within the worldline formalism, that the box diagram (see the left diagram in Fig.~\ref{fig2}) contains a pole $1/l^2$ if it is calculated in off-forward kinematics. This may be viewed as a nonlocal generalization of the local relation~(\ref{trian}) unintegrated in the Bjorken variable $x$. As envisaged in~\cite{Jaffe:1989jz} and elaborated in~\cite{Tarasov:2020cwl}, at least after the $x$-integration, the pole should be canceled by another massless pole due to the exchange of the $\eta_0$ meson, the would-be Nambu-Goldstone boson of $U_A(1)$ symmetry breaking. This requirement leads to an independent derivation of the $U_A(1)$ Goldberger-Treiman relation~\cite{Hatsuda:1989bi,Veneziano:1989ei,Shore:2007yn} between the pseudoscalar and pseudovector form factors.

Motivated by these developments, in this paper we further explore the physics of anomaly poles in two different directions. First, we calculate the box diagram in off-forward kinematics in the standard perturbation theory. This is a useful cross-check of the result obtained in the worldline formalism~\cite{Tarasov:2021yll}. In addition to reproducing the pole term, we  obtain the `usual' perturbative corrections to the $g_1(x)$ structure function  which features the DGLAP splitting function and a coefficient function. We then interpret the result in terms of the generalized parton distributions (GPDs) $\tilde{H}$ and $\tilde{E}$. The emergence of the pole is potentially problematic for the QCD factorization of the Compton amplitude. 
We discuss how factorization may still be justified following the possibility of cancellation of poles.

Second, we point out that  entirely analogous poles can arise in unpolarized DIS, or more precisely, in the symmetric (in Lorentz indices $\mu\nu$) part of the Compton scattering amplitude $T^{\mu\nu}$ in off-forward kinematics. Just as the pole in the polarized sector is related to the axial (chiral) anomaly, that in the unpolarized sector is  related to the trace anomaly. Indeed, it is known in QED and other gauge theories~\cite{Giannotti:2008cv,Armillis:2009im,Donoghue:2015xla} that the off-forward photon matrix element of the energy momentum tensor $\Theta^{\mu\nu}$ has an anomaly pole:
\beq
\langle p_2|\Theta^{\mu\nu}|p_1\rangle \sim \frac{1}{l^2}\langle p_2|F^{\alpha\beta}F_{\alpha\beta}|p_1\rangle \, , \label{traceano}
\eeq
again from the triangle diagram. 
The residue is proportional to the matrix element of the twist-four scalar operator $\langle F^{\alpha\beta}F_{\alpha\beta}\rangle$ (or the `gluon condensate' in QCD) which characterizes the trace anomaly. We shall derive the unintegrated (in $x$) version of~(\ref{traceano}) by evaluating the quark box diagrams and interpret the result in terms of the unpolarized GPDs $H$ and $E$. We then make a connection to the gravitational form factors of the proton and discuss the possibility of cancellation of poles.

\section{Preliminaries}
\begin{figure}[t]
\centering
\includegraphics[width = 17cm]{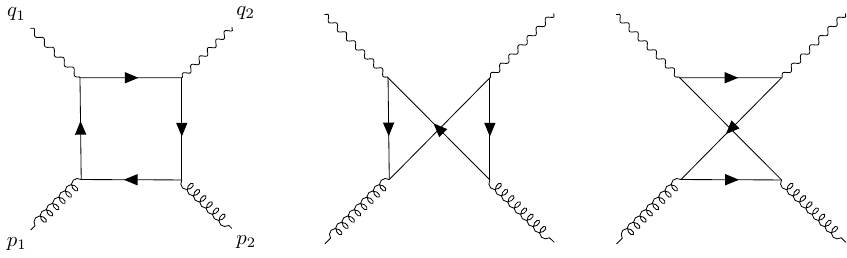}
\caption{Box diagrams for the Compton amplitude in  off-forward kinematics.}
\label{fig2}
\end{figure}

In this section, we set up our notations for the kinematical variables that enter the calculation of the quark box diagrams in  DIS.  More precisely,  since we generalize the calculation to off-forward kinematics as explained in the Introduction, we consider  the Compton scattering amplitude 
\beq
T^{\mu\nu}=i\int \frac{d^4y}{2\pi}e^{iq\cdot y}\langle P_2|{\rm T}\{J^\mu(y/2)J^\nu(-y/2)\}|P_1\rangle =T^{\mu\nu}_{\rm sym} + iT^{\mu\nu}_{\rm asym} \, , \label{comp}
\eeq
where $J^\mu=\sum_f e_f \bar{\psi}_f\gamma^\mu \psi_f$, with  $f=u,d,s,..$ being a flavor index, is the electromagnetic current and the subscript `sym/asym' refers to the symmetric/antisymmetric part in the photon polarization indices $\mu,\nu$. $|P_{1,2}\rangle$ are the proton single-particle  states.  $q=\frac{q_1+q_2}{2}$ is the average of the incoming and outgoing virtual photon momenta, and the momentum transfer is denoted by  $P_2 -P_1 =q_1-q_2= l$. 
 We assume $-q^2\equiv Q^2>0$ is large and neglect  `higher-twist' terms of order ${\cal O}(M^2/Q^2)$  where  $M^2=P_1^2=P_2^2$ is the proton mass squared. Their inclusion is understood in the literature \cite{Braun:2014sta}. We also assume that $t\equiv l^2<0$ is much smaller than the hard scale $|t|\ll Q^2$ and neglect terms of order $l^2/Q^2$.
We define the following  variables
\beq
&&P=\frac{P_1+P_2}{2},  \qquad 
x_B=\frac{Q^2}{2P\cdot q}, \qquad  
\xi=\frac{q_2^2-q_1^2}{4P\cdot q}\approx \frac{-l^+}{2P^+} \, ,
\eeq
where $x_B$ is the generalized Bjorken variable and $\xi$ is the skewness parameter. In forward scattering, $x_B$ coincides with the usual Bjorken variable in DIS. In Deeply Virtual Compton Scattering (DVCS) where $q_2^2=0$, $x_B\approx \xi$.

The quark box diagrams of interest are part of the  perturbative expansion of~(\ref{comp}) at one-loop. There are three topologies, see Fig.~\ref{fig2}. The diagrams consist of two insertions of photon fields with  momenta ($q_{1},q_{2}$) and two insertions of gluon fields with partonic momenta  ($p_{1},p_{2}$) which we parametrize as 
\begin{equation}
p_1 = p - \dfrac{l}{2} \, , \qquad 
p_2= p+\dfrac{l}{2} \, , \qquad x\equiv \frac{p\cdot q}{P\cdot q}\, .
\end{equation}
Note that $p_2-p_1=P_2-P_1=l$, so the momentum transfer $t=l^2<0$ is the same in both the hadronic and partonic processes. 
We  assume the incoming partons to be massless, $p_1^2=p_2^2=0$, which means $p\cdot l=0$ and $p^2=-l^2/4$. 
The kinematical variables at the partonic level are defined as
\beq
\hat{x}=\frac{Q^{2}}{2p \cdot q}=\frac{x_B}{x}, 
\qquad \hat{\xi}= \frac{q_2^2-q_1^2}{4p\cdot q} = \frac{-q\cdot l}{2p\cdot q} = \frac{\xi}{x}\,.
\eeq
The relation $\hat{x} q\cdot l=-\hat{\xi}Q^2$  will be used in the calculation below. The photon virtualities can be written as
\beq
q_1^2= -Q^2\frac{\hat{x}+\hat{\xi}}{\hat{x}} +\frac{l^2}{4}, \qquad q_2^2 = Q^2\frac{\hat{\xi}-\hat{x}}{\hat{x}} + \frac{l^2}{4} \, . \label{q12}
\eeq
Finally, the  polarization vectors of the incoming and outgoing gluons, $\epsilon(p_1)\equiv \epsilon_1$, $\epsilon^*(p_2)\equiv \epsilon^*_2$, respectively, satisfy the physical conditions  
\begin{align}
\epsilon_1\cdot (p-l/2)=0 \quad & \to \quad \epsilon_1 \cdot p = \frac{\epsilon_1\cdot l}{2} \, , 
\nn
\epsilon_2^*\cdot (p+l/2)=0 \quad & \to \quad \epsilon_2^*\cdot p = -\frac{\epsilon_2^*\cdot l}{2} \, . \label{epscond}
\end{align}
We note that throughout this paper we use the conventions $\gamma_5=i\gamma^0\gamma^1\gamma^2\gamma^3$ and $\epsilon^{0123}=+1$. 
This becomes relevant for the antisymmetric part of the Compton amplitude~(\ref{comp}), to which we turn first.

\section{Antisymmetric part}
In the antisymmetric case we define 
\beq
{\cal J}^\alpha&\equiv&-\epsilon^{\alpha\beta\mu\nu}P_\beta {\rm Im}T^{\rm asym}_{\mu\nu} \, .
\eeq
In the forward limit $l^\mu \to 0$, ${\cal J}^\alpha=g_1(x_B)\bar{u}(P)\gamma^\alpha\gamma_5 u(P)=2g_1(x_B)S^\alpha$ is proportional to the $g_1$ structure function in polarized DIS. 
We have calculated the box diagrams in the near-forward region $|l^\mu|\ll Q$, using the Mathematica package `Package-X'~\cite{Patel:2015tea}. The result is, for massless quarks in the loop,  
\beq
\left.{\cal J}^\alpha\right|_{\rm box}\approx \frac{1}{2} \frac{\alpha_s}{2\pi}\left(\sum_f e_f^2\right)  \bar{u}(P_2)\left[\left(\Delta P_{qg}\ln\frac{Q^2}{-l^2}+\delta C_g^{\rm off}\right)\otimes \Delta G(x_B) \gamma^\alpha\gamma_5  +\frac{l^\alpha}{l^2} \delta C_g^{\rm anom}\otimes \tilde{{\cal F}}(x_B)\gamma_5\right]u(P_1)\, , \label{off}
\eeq
where $A\otimes B(x_B) \equiv \int_{x_B}^1 \frac{dx}{x} A(\frac{x_B}{x})B(x)$. In~(\ref{off}), we have neglected all terms that vanish in the limit $l\to 0$, including the $\xi$-dependence of various coefficients. (The dependence on $\xi$ will be  kept in the next section.) 
$\Delta P_{qg}(\hat{x})=2T_R(2\hat{x}-1)$ is the polarized $g\to q$ splitting function with $T_R=\frac{1}{2}$ being the color factor, and 
\beq
\delta C_g^{\rm off}(\hat{x}) = 2T_R(2\hat{x}-1)\left( \ln \frac{1}{\hat{x}(1-\hat{x})}-1\right)
\eeq
is the coefficient function. (In both $\Delta P_{qg}$ and $\delta C_g^{\rm off}$ the explicit factor of 2 accounts for the quark and antiquark contributions.) 
$\tilde{{\cal F}}(x)$ is 
the twist-four pseudoscalar parton distribution~\cite{Tarasov:2020cwl,Hatta:2020ltd,Radyushkin:2022qvt}
\beq
\tilde{{\cal F}}(x,l^2)&\equiv & \frac{iP^+}{\bar{u}(P_2)\gamma_5u(P_1)}\int \frac{dz^-}{2\pi} e^{ixP^+z^-} \langle P_2|F_a^{\mu\nu}(-z^-/2)\tilde{F}^a_{\mu\nu}(z^-/2)|P_1\rangle \, ,
\eeq
entering Eq.~(\ref{off}) with 
\beq 
\delta C_g^{\rm anom}(\hat{x}) =4T_R(1-\hat{x}) \, .
\eeq
We note that in the one-loop calculation, $\Delta G(x)$ and $\tilde{{\cal F}}(x)$ have been identified via the tensor structures 
$\frac{-i}{p^+}\langle p_2|F^{+\mu}\tilde{F}^+_{\ \mu}|p_1\rangle \propto i\epsilon^{+\nu\alpha\beta}p_\nu\epsilon^*_{2\alpha}\epsilon_{1\beta} \equiv i\epsilon^{+p\epsilon_2^*\epsilon_1}$ and 
$\langle p_2|F^{\mu\nu}\tilde{F}_{\mu\nu}|p_1\rangle\propto \epsilon^{\mu\nu\alpha\beta}l_\mu p_\nu \epsilon_{1\alpha}\epsilon^*_{2\beta}\equiv \epsilon^{lp\epsilon_1 \epsilon^*_2}$, respectively. 

The pole term $1/l^2$ in~(\ref{off}) agrees with the result derived in Ref.~\cite{Tarasov:2020cwl}  using the worldline formalism. In the Appendix A, we also rederive this result via an ordinary Feynman diagram calculation.  Although not manifest, the numerator actually contains an extra factor of $l$, so this term is singular in the sense that the limit $\lim_{l\to 0}l^\alpha l^\beta/l^2$ depends on the direction of $l^\mu$.  In addition, we find the expected contribution to the $g_1$ structure function where $l^2$ plays the role of an infrared cutoff. Both $\Delta P_{qg}$ and $\delta C_g^{\rm off}$ satisfy $\int_0^1 d\hat{x} \Delta P_{qg}(\hat{x})=\int_0^1d\hat{x} \delta C_g^{\rm off}(\hat{x})=0$. They thus do not contribute to the first moment of $g_1$. This is a necessary condition in order to be consistent with the perturbative result~(\ref{trian}) for the local operator $J_5^\mu$.  Note that in dimensional regularization in $4-2\epsilon$ dimensions, one has instead 
\beq
\Delta P_{qg}\frac{-1}{\epsilon} + \delta C_g^{\overline{\rm MS}}, \qquad \delta C_g^{\overline{\rm MS}}(\hat{x})=2T_R(2\hat{x}-1)\left(\ln\frac{1-\hat{x}}{\hat{x}}-1\right) + 4T_R(1-\hat{x}) \, , 
\eeq
in the coefficient of $\Delta G(x_B)$. 
As pointed out in~\cite{Vogelsang:1990ug}, the last term $4T_R(1-\hat{x})$ is of infrared origin. In  off-forward regularization, this term becomes the pole term.

The cancellation  of the pole in~(\ref{off}) and its implications have been discussed in~\cite{Tarasov:2021yll}. Here we point out that the description of the pole is most naturally phrased  in terms of the generalized parton distributions (GPDs). For this purpose we return to~(\ref{comp}) and fully treat it as an off-forward amplitude.  In the GPD factorization framework~\cite{Collins:1998be,Ji:1998xh}, the imaginary part of the Compton amplitude can be expanded as 
\beq
{\cal J}^\alpha &=&\frac{1}{2}\sum_f e_f^2\bar{u}(P_2)\left[\gamma^\alpha\gamma_5 (\tilde{H}_f(x_B,\xi,l^2)+\tilde{H}_f(-x_B,\xi,l^2)) +\frac{l^\alpha\gamma_5}{2M}(\tilde{E}^{\rm bare}_f(x_B,\xi,l^2)+\tilde{E}^{\rm bare}_f(-x_B,\xi,l^2)) \right]u(P_1) \nn
&& +{\cal O}(\alpha_s)+{\cal O}(1/Q^2) \, ,  \label{expand}
\eeq 
where $\tilde{H}_f$ and $\tilde{E}_f$ are the  polarized quark GPDs 
for flavor $f$. The meaning of the superscript `bare' will be explained shortly. 
In the forward limit, the polarized quark and antiquark PDFs are recovered, $\tilde{H}_q(x_B)=\Delta q(x_B)$ and  $\tilde{H}_q(-x_B)=\Delta \bar{q}(x_B)$, so that ${\cal J}^\alpha =\sum_f e_f^2 (\Delta q_f(x_B)+\Delta \bar{q}_f(x_B)) S^\alpha+\cdots \approx 2g_1(x_B)S^\alpha$. 
It is tempting to regard~(\ref{off}) as part of the ${\cal O}(\alpha_s)$ corrections in~(\ref{expand}). However, 
clearly the pole term $1/l^2$  does not fit into the general structure of factorization. This term remained unnoticed in the calculation of the antisymmetric part of the one-loop corrections to the Compton amplitude~\cite{Belitsky:1997rh,Ji:1998xh}. The reason is that in the GPD-based calculations in the literature, one sets $l^\mu \approx -2\hat{\xi} p^\mu$ {\it before} the loop integration. Since $l^2=0$ in this approximation,  the pole $1/l^2$ cannot arise. Without this regulator,  the box diagrams are power divergent in the infrared  unless some cancellation mechanism is at work. In the present case, the matrix element of the twist-four operator 
$\langle p_2|F^{\mu\nu}\tilde{F}_{\mu\nu}|p_1\rangle\propto \epsilon^{\mu\nu\alpha\beta}l_\mu p_\nu \epsilon_{1\alpha}\epsilon^*_{2\beta}$ vanishes if one assumes $l^\mu \propto p^\mu$.{\footnote{Introducing the current quark mass $m_q$ may regularize the pole $1/l^2$, but only when $\sqrt{|l^2|}\sim m_q$, so the problem persists. We leave the calculation with finite quark mass to future work. }


 The problem is that,  despite the proportionality to the twist-four distribution, the pole term in (\ref{off}) cannot be dismissed as a higher twist contribution because there is no suppression factor $1/Q^2$. In fact, it is parametrically of the same order as the twist-two $\Delta G(x)$ term in (\ref{off})  since $l^\alpha \langle F\tilde{F}\rangle/l^2\sim  l^\alpha l^\beta/l^2\sim {\cal O}(1)$ as mentioned above.  
Note also that the pole term has the same Lorentz structure $\propto l^\alpha$ as the GPD $\tilde{E}$ in~(\ref{expand}). 
Clearly, this poses a threat to QCD factorization for the Compton amplitude in the GPD $\tilde{E}$ sector.\footnote{Note that the experimentally accessible Compton amplitude corresponds to $q_2^2\ge 0$ or $\xi\gtrsim x_B$, whereas  we approximated $\xi\approx 0$ in~(\ref{off}). However, we have checked that the pole $1/l^2$ exists also for $\xi>0$. Besides, the general argument for factorization does not rely on the relative magnitude of  $x_B$ and $\xi$.}

On the other hand, the Compton amplitude  must have a well-defined forward limit. Thus, there must be a mechanism to cancel the pole at $l^2=0$ which could ultimately `rescue' the factorization theorem.  Noticing that the pole term comes entirely from the infrared region of the box diagram, we absorb it into the definition of the GPD $\tilde{E}$ and let it cancel with a `primordial' pole  in $\tilde{E}^{\rm bare}$. Namely, we redefine as 
\beq
\tilde{E}_f(x_B,l^2) +\tilde{E}_{f}(-x_B,l^2) =\tilde{E}^{\rm bare}_f(x_B,l^2) +\tilde{E}^{\rm bare}_{f}(-x_B,l^2)  +  \frac{\alpha_s}{2\pi} \frac{2M}{l^2}\delta C_g^{\rm anom}\otimes \tilde{{\cal F}}(x_B,l^2) \, , \label{canc}
\eeq
and postulate that $\tilde{E}^{\rm bare}$ is dominated by a pole  
\beq
\tilde{E}^{\rm bare}_f(x_B,l^2) +\tilde{E}^{\rm bare}_{f}(-x_B,l^2) \approx - \frac{\alpha_s}{2\pi} \frac{2M}{l^2}\delta C_g^{\rm anom}\otimes \tilde{{\cal F}}(x_B,l^2=0) \, , \label{rest}
\eeq
which exactly cancels the perturbative pole in the last term. We further postulate that it is this `renormalized' $\tilde{E}$ that integrates to the pseudoscalar form factor. Namely, 
\begin{align}
g_A(l^2) & =\sum_f \int_{-1}^1 dx \tilde{H}_f(x,\xi,l^2) = \sum_f\int_0^1 dx (\tilde{H}_f(x,\xi,l^2)+\tilde{H}_{f}(-x,\xi,l^2)) \, , \\ 
g_P(l^2) & =\sum_f\int_{-1}^1 dx \tilde{E}_f(x,\xi,l^2) =\sum_f \int_0^1 dx (\tilde{E}_f(x,\xi,l^2)+\tilde{E}_{f}(-x,\xi,l^2)) \, ,  \label{gpint}
\end{align}
where $g_A$ and $g_P$ are the singlet axial form factors defined by 
\beq
\langle P_2|J_5^\alpha |P_1\rangle
= \bar{u}(P_2)\left[\gamma^\alpha\gamma_5 g_A(l^2) +\frac{l^\alpha \gamma_5}{2M}g_P(l^2)\right]u(P_1) \, .
\label{factors}
\eeq
The prescription~(\ref{canc}) is outside the conventional  logarithmic renormalization of twist-two GPDs. In order to better justify it, we show in the Appendix (see~(\ref{polepole})) that exactly the same pole structure arises in the one-loop calculation of the polarized quark  GPD of a gluon\footnote{We thank Swagato Mukherjee for suggesting this calculation.}:
\beq
p^+\int \frac{dz^-}{4\pi}e^{i\hat{x}p^+z^-}\langle p_2|\bar{\psi}(-z^-/2)\gamma^+\gamma_5\psi(z^-/2)|p_1\rangle\biggl|_{\rm pole}
\sim  \frac{\alpha_s}{2\pi}T_R  \frac{2il^+}{l^2}(1-\hat{x})\otimes \delta(1-\hat{x})\epsilon^{\epsilon_1 \epsilon_2^*lp} \, .
\eeq 
Moreover, performing the $x$-integral in~(\ref{gpint}), we obtain   
\beq
\frac{g_P(l^2)}{2M}= -\frac{i}{l^2} \left(\left.\frac{\langle P_2|\frac{n_f\alpha_s}{4\pi} F\tilde{F}|P_1\rangle}{\bar{u}(P_2)\gamma_5u(P_1)}\right|_{l^2=0} - \frac{\langle P_2|\frac{n_f\alpha_s}{4\pi} F\tilde{F}|P_1\rangle}{\bar{u}(P_2)\gamma_5u(P_1)} \right)\, .
\label{cf2}
\eeq
By construction, this is finite at $l^2=0$. One can check that (\ref{cf2}) is consistent with the following exact relation obtained by taking the divergence of~(\ref{factors}):
\beq
2Mg_A(l^2) + \frac{l^2g_P(l^2) }{2M}
=i\frac{\langle P_2|\frac{n_f\alpha_s}{4\pi} F\tilde{F}|P_1\rangle}{\bar{u}(P_2)\gamma_5u(P_1)} \, , \label{exact2}
\eeq 
under the assumption that $g_A(l^2)\approx g_A(0)=\Delta\Sigma$ slowly varies with $l^2$. These  arguments suggest that the pole should indeed be regarded as a part of the quark GPD $\tilde{E}$, although a more rigorous, field-theoretic justification of~(\ref{canc}) is needed.

The cancellation of poles within the form factor $g_P$ is the central argument of~\cite{Tarasov:2021yll} following the earlier suggestion in~\cite{Jaffe:1989jz}. The  massless  pole in the first term of~(\ref{cf2})  can be  nonperturbatively generated by the $t$-channel exchange of the would-be Nambu-Goldstone boson of spontaneous $U_A(1)$ symmetry breaking, the primordial $\eta_0$ meson. This is analogous to the pion pole in the $SU_A(2)$ pseudoscalar form factor $g_P^{(3)} \sim  \frac{1}{l^2-m_\pi^2}$. The difference, however, is that the latter is physical while the former is not. In reality, the $U_A(1)$ symmetry is explicitly broken by the anomaly. In the present context, this effect is represented by the second term in~(\ref{cf2}), turning the massless pole into a massive one at the physical $\eta'$ meson mass $m_\eta^2$. Indeed, the form factor $\langle P_2|F\tilde{F}|P_1\rangle$ has a pole at $l^2=m_{\eta'}^2$, and in the single-pole approximation~(\ref{exact2}) is satisfied by 
\beq
\frac{g_P(l^2)}{2M} \approx  \frac{-2M\Delta \Sigma}{l^2-m_{\eta'}^2}\,, \qquad  i\frac{\langle P_2|\frac{n_f\alpha_s}{4\pi} F\tilde{F}|P_1\rangle}{\bar{u}(P_2)\gamma_5u(P_1)} \approx -2M\Delta\Sigma\frac{m_{\eta'}^2}{l^2-m_{\eta'}^2} \, .
\eeq

Returning to GPDs, we add that, in order to guarantee the finiteness of the Compton amplitude, there must be a massless pole already in the GPD $\tilde{E}$.\footnote{The pion pole is argued to exist also in the GPD $\tilde{E}$ in the isovector channel  $\tilde{E}_u-\tilde{E}_d \sim \frac{1}{l^2-m_\pi^2}$~\cite{Penttinen:1999th}.} Moreover, the cancellation of poles in~(\ref{canc}) must somehow occur  for all values of $x_B$ and $Q^2$ (the latter enters as the renormalization scale), or else the factorization of the Compton amplitude is in danger. On the other hand, 
since the cancellation of poles occurs within the GPD $\tilde{E}$ sector alone, we do not see any issues with the usual factorization of the $g_1$ structure function in polarized DIS.
In this sense, the above off-forward calculation may be regarded as an alternative to other regularization schemes  such as the $\overline{\rm MS}$ scheme.  

\section{Symmetric part}
Remarkably, the $1/l^2$ pole is also present in the symmetric (in $\mu\nu$) part of the Compton amplitude~(\ref{comp}). As mentioned in the Introduction, and will be studied in detail below, in this case the pole is related to the QCD trace anomaly. In this section we obtain the fully analytic expression of the pole terms and discuss their implications for the Compton amplitude and the gravitational form factors.

Let us first recall that in the forward limit, the imaginary part of the symmetric Compton amplitude $T_{\rm sym}^{\mu\nu}$ is related to the $F_1$ and $F_2$ structure functions,
\beq
{\rm Im} T^{\mu\nu}_{\rm sym}= 
\left(-g^{\mu\nu}+\frac{q^\mu q^\nu}{q^2}\right) F_1 (x_B)+ \left(P^\mu-\frac{P\cdot q}{q^2}q^\mu\right)\left(P^\nu-\frac{P\cdot q}{q^2}q^\nu\right)\frac{2x_B F_2(x_B)}{Q^2} \, .
\eeq
To leading order, 
\beq
F_2(x)= \sum_f e_f^2x\big(q_f(x)+\bar{q}_f(x)\big) =2xF_1(x) \, ,
\eeq
where $q_f(x)$ ($\bar{q}_f(x))$) is the unpolarized quark (antiquark) PDF.
The first moment of $F_2$ is related to the matrix element of the quark part of the QCD energy momentum tensor $\Theta^{\mu\nu}_f=\bar{q}_f\gamma^{(\mu}iD^{\nu)}q_f$:
\beq 
\langle P|\Theta^{++}_f|P\rangle = 2 (P^+)^2 A_f \, , \qquad A_f=\int_0^1 dx x \big(q_f(x)+\bar{q}_f(x)\big) \, . \label{def}
\eeq
Physically, $A_f$ is the momentum fraction of the proton carried by $f$-quarks and antiquarks. 

Turning to the off-forward case, we have calculated the contribution from the box diagrams using Package-X~\cite{Patel:2015tea} and found a  more complicated tensor structure,
\beq
\left.{\rm Im} T^{\mu\nu}_{\rm sym}\right|_{\rm box}= 
\left(-g^{\mu\nu}+\frac{q^\mu q^\nu}{q^2}\right) F^{\rm off}_1(x_B,l) + \left(P^\mu-\frac{P\cdot q}{q^2}q^\mu\right)\left(P^\nu-\frac{P\cdot q}{q^2}q^\nu\right)\frac{2x_BF^{\rm off}_2(x_B,l)}{Q^2} 
+\dots,
\label{offf}
\eeq
where the ellipses denote terms proportional to tensors like  $l^{(\mu} P^{\nu)}$, $l^{(\mu} q^{\nu)}$ and $l^\mu l^\nu$ such that they satisfy the conservation law $q_\mu {\rm Im}T^{\mu\nu}_{\rm sym}=0$. This time $F_{1,2}^{\rm off}$ are related to the unpolarized GPDs instead of the PDFs. Different from the polarized case, for a reason to become clear shortly, here we shall keep the full dependence on the skewness parameter $\xi$. For this purpose, we will make the replacement $l^\mu \approx  -2\xi P^\mu \approx - 2\hat{\xi}p^\mu$ in the tensors $l^{(\mu} P^{\nu)}$, $l^{(\mu} q^{\nu)}$ and $l^\mu l^\nu$, but {\it  not} in potential pole terms $\propto 1/l^2$. 
In doing so, we find that Eq.~(\ref{offf}) takes the simple form
\beq
\left.{\rm Im} T^{\mu\nu}_{\rm sym}\right|_{\rm box} \approx 
\left(-g^{\mu\nu}+\frac{q^\mu q^\nu}{q^2}\right) \bar{F}^{\rm off}_1(x_B,l) + \left(P^\mu-\frac{P\cdot q}{q^2}q^\mu\right)\left(P^\nu-\frac{P\cdot q}{q^2}q^\nu\right)\frac{2x_B\bar{F}^{\rm off}_2(x_B,l)}{Q^2} \, ,
\eeq
where (setting $\xi=0$ in non-pole terms)
\beq
\begin{split}
&\bar{F}^{\rm off}_1(x_B,l) \approx \frac{1}{2} \frac{\alpha_s}{2\pi}\left(\sum_f e_f^2\right)  \left[ \left(P_{qg}\ln \frac{Q^2}{-l^2}+C^{\rm off}_{1g}\right) \otimes g(x_B) 
+ \frac{1}{l^2}C^{\rm anom} \otimes' {\cal F} (x_B,\xi,l^2)\frac{\bar{u}(P_2)u(P_1)}{2M}\right] \, , \\
& \bar{F}^{\rm off}_2(x_B,l) \approx   x_B\frac{\alpha_s}{2\pi}\left(\sum_f e_f^2\right)  \left[\left( P_{qg}\ln \frac{Q^2}{-l^2}+C^{\rm off}_{2g}\right) \otimes g(x_B) + \frac{1}{l^2}C^{\rm anom}\otimes' {\cal F} (x_B,\xi,l^2)\frac{\bar{u}(P_2)u(P_1)}{2M}\right] \, .
\label{offsym} 
\end{split}
\eeq
We recognize the expected structure of the one-loop corrections associated with the unpolarized gluon PDF $g(x)$, with the splitting function  $P_{qg}(\hat{x})=2T_R((1-\hat{x})^2+\hat{x}^2)$. The coefficient functions  are given by 
\beq
\begin{split}
& C_{1g}^{\rm off}(\hat{x}) =2T_R ((1-\hat{x})^2+\hat{x}^2) \left(\ln \frac{1}{\hat{x}(1-\hat{x})}-1\right) \, , \label{coeff} \\
& C_{2g}^{\rm off}(\hat{x})=2T_R((1-\hat{x})^2+\hat{x}^2) \left(\ln \frac{1}{\hat{x}(1-\hat{x})}-1\right) +8T_R\hat{x}(1-\hat{x}) \, .
\end{split}
\eeq
In addition, we find a pole $1/l^2$ in both $\bar{F}_1^{\rm off}$ and $\bar{F}_2^{\rm off}$ (but not in the difference $\bar{F}^{\rm off}_2-2x_B \bar{F}^{\rm off}_1$ relevant to the longitudinal structure function), with the following convolution formula 
\beq
C^{\rm anom}\otimes' {\cal F} (x_B,\xi,l^2) \equiv \int_{x_B}^1\frac{dx}{x}K(\hat{x},\hat{\xi}){\cal F} (x,\xi,l^2)-\frac{\theta(\xi-x_B)}{2}\int_{-1}^1 \frac{dx}{x}L(\hat{x},\hat{\xi}){\cal F} (x,\xi,l^2) \, , \label{two}
\eeq
where 
\beq
&&K (\hat{x},\hat{\xi})  
=2T_R\frac{\hat{x}(1-\hat{x})}{1-\hat{\xi}^2} \, , \qquad L(\hat{x},\hat{\xi}) = 2T_R\frac{\hat{x}(\hat{\xi}-\hat{x})}{1-\hat{\xi}^2} \, . \label{whitec}
\eeq
The twist-four scalar gluon  GPD is defined as 
\beq
{\cal F} (x,\xi,l^2)&=&-4xP^+M \int \frac{dz^-}{2\pi} e^{ixP^+z^-} \frac{ \langle P_2|F^{\mu\nu}(-z^-/2)F_{\mu\nu}(z^-/2)|P_1\rangle}{\bar{u}(P_2)u(P_1)} \, . \label{four}
\eeq
Several comments are in order. First, the two terms in~(\ref{two}) come from the first and third diagrams in Fig.~\ref{fig2}, respectively. The 
imaginary part of the latter is nonvanishing only in the region  $0<x_B<\xi$ where the outgoing virtual photon becomes timelike, $q_2^2>0$, see~(\ref{q12}). The structure of the convolution~(\ref{two}) has been  inferred from the analysis of Compton scattering in a related context~\cite{White:2001pu}. 
Second, concerning the twist-four gluon GPD~(\ref{four}),  the original definition in~\cite{Hatta:2020iin} was a forward matrix element $\langle P|...|P\rangle$. The necessity to define ${\cal F}$ in off-forward kinematics was later emphasized in~\cite{Radyushkin:2021fel}, and this turns out to be  relevant to the present discussion. Accordingly, we have rescaled the distribution by $\bar{u}(P_2)u(P_1)$.
We have also introduced the prefactor $x$ to compensate for the $1/x$ factor in (\ref{two}). This factor makes ${\cal F}(x,\xi,l^2)$ an odd function in $x$, just like $g(x)$.  
 Finally, in the actual one-loop calculation the gluon PDF can be identified as $g(x)\sim \langle -F^{+\mu}F^+_{\ \mu}\rangle \propto \epsilon_1\cdot \epsilon^*_2$, while the pole term $\frac{1}{l^2}{\cal F}(x,\xi,l^2)$ comes from  the structure\footnote{Actually the box diagram  generates  an additional tensor structure 
\beq
 \epsilon_{2}^{*\mu} \epsilon_1^\nu + \epsilon_{2}^{*\nu}\epsilon_1^\mu -g^{\mu\nu}_\perp \epsilon_2^*\cdot \epsilon_1 \, ,  \label{transv}
\eeq  
and this has been  taken into account when determining the coefficient functions~(\ref{coeff}). 
For spin-1 hadrons,~(\ref{transv}) induces an additional structure function in~(\ref{offf})~\cite{Jaffe:1989xy}.
For spin-$\frac{1}{2}$ hadrons, the structure~(\ref{transv}) is neglected in inclusive DIS because it vanishes after  averaging over the gluon polarizations~(\ref{assumefact}). In Compton scattering, it becomes part of the gluon transversity GPD~\cite{Hoodbhoy:1998vm}.} 
\beq
\frac{1}{l^2}\langle p_2,\epsilon_2|F^{\mu\nu}F_{\mu\nu}|p_1,\epsilon_1\rangle \propto \frac{1}{l^2}\left(-\epsilon_1\cdot p_2 \epsilon_2^* \cdot p_1 +p_1\cdot p_2\epsilon_1\cdot \epsilon_2^*\right)  = \frac{1}{l^2}\left(\epsilon_1\cdot l \epsilon_2^* \cdot l - \frac{l^2}{2}\epsilon_1\cdot \epsilon_2^*\right)  \label{onshell}
\,.
\eeq
 This is again order unity $l^\alpha l^\beta/l^2\sim {\cal O}(1)$, the same order as the twist-two $g(x)$ term in (\ref{offsym}). 
Equation~(\ref{onshell})  shows that, for on-shell gluon states, the tree-level matrix element $\langle F^{\mu\nu}F_{\mu\nu}\rangle$ is nonvanishing only in  off-forward kinematics $l^\mu\neq 0$. If one averages over the two gluon polarization states by making the replacement 
\beq
\epsilon_{2\alpha}^*\epsilon_{1\beta} \to -\dfrac{1}{2} g_{\alpha\beta}^\perp  \, ,
\label{assumefact}
\eeq
where $g^\perp_{\alpha \beta} \equiv g_{\alpha \beta} - n_{\alpha} \bar{n}_\beta - \bar{n}_{\alpha} n_{\beta}$ with the light-like vectors $n = (1,0,0,-1)/\sqrt{2}$ and $\bar{n} = (1,0,0,1)/\sqrt{2}$, the pole disappears because
\beq
\frac{1}{l^2}\left(\epsilon_1\cdot l\; \epsilon_2^* \cdot l - \frac{l^2}{2}\epsilon_1\cdot \epsilon_2^*\right) \to \frac{1}{2l^2} (\vec{l}_\perp^{\;2} + l^2) = \dfrac{\hat{\xi}^2}{2} \, . \label{nopole}
\eeq
Again, this is the reason why the pole has remained unnoticed in the one-loop GPD calculations in the unpolarized sector~\cite{Ji:1997nk,Mankiewicz:1997bk,Belitsky:1997rh,Pire:2011st} (see also a recent two-loop calculation in DVCS~\cite{Braun:2022bpn}) where one projects onto the twist-two gluon GPDs $H_g$ and $E_g$ via~(\ref{assumefact}). Without this projection, the loop integral is power divergent if one sets $l^\mu=-2\hat{\xi}p^\mu$ from the outset.  

While the replacement~(\ref{assumefact}) may be justified for the unpolarized DIS structure functions where the initial and final proton spins are equal and averaged over as 
$\sum_S \langle PS|...|PS\rangle$, this is not the case for GPDs $\langle P_2S_2|...|P_1S_1\rangle$ in Compton scattering where the initial and final spin states $S_1$ and $S_2$ are typically unconstrained. Moreover, in contrast to the partonic matrix element $ \langle p_2|F^{\mu\nu}F_{\mu\nu}|p_1\rangle$ (see~(\ref{onshell})), the nonperturbative {\it proton} matrix element $\langle P_2|F^{\mu\nu}F_{\mu\nu}|P_1\rangle$ does not vanish in the forward limit, not even after averaging over the proton spin states. In particular, the zeroth moment of ${\cal F}(x,\xi=0,l^2=0)$ is the so-called gluon condensate:
\beq
\int \frac{dx}{x}  {\cal F}(x,0,0) \propto \langle P|F^{\mu\nu}F_{\mu\nu}|P\rangle \, , \label{47}
\eeq
responsible for the generation of the proton mass via the QCD trace anomaly. As a matter of fact, the connection between trace anomalies and massless poles $1/l^2$ is known in the literature~\cite{Giannotti:2008cv,Armillis:2009im,Donoghue:2015xla}. An explicit calculation of the triangle diagram shows that the matrix element of the QED energy momentum tensor between on-shell photon states has a pole:
\beq
\langle p_2|\Theta_{\rm QED}^{\mu\nu}|p_1\rangle = -\frac{ e^2}{24\pi^2 l^2} \left(p^\mu p^\nu + \frac{l^\mu l^\nu-l^2 g^{\mu\nu}}{4}\right)\langle p_2|F^{\alpha\beta}F_{\alpha\beta}|p_1\rangle +\ldots \, . \label{qed}
\eeq
If one introduces the usual color factor $T_R=\frac{1}{2}$ on the right-hand side, essentially the same formula holds for the one-loop gluon matrix element of the quark energy momentum tensor $\Theta^{\mu\nu}_f$ in QCD. 
Taking the $++$ component, one finds a relation 
\beq
\langle p_2|\Theta_f^{++}|p_1\rangle \sim -\frac{T_R\alpha_s\,(p^+)^2}{6\pi l^2} (1+\hat{\xi}^2)\langle p_2|F^2|p_1\rangle \, . \label{af}
\eeq 
On the other hand, if one takes the trace of~(\ref{qed}), the pole disappears. Applied to the QCD case we find
\beq
\langle p_2|(\Theta_f)^\mu_{\mu}|p_1\rangle  = \langle p_2| \frac{T_R\alpha_s }{6\pi}F^2|p_1\rangle \, ,
\eeq
and the right-hand side is nothing but the contribution to the QCD trace anomaly from the  single-flavor quark energy momentum tensor to one-loop\footnote{The separation of the total trace $\Theta^\mu_\mu$ into the quark and gluon parts $(\Theta_f)^\mu_\mu$ and $(\Theta_g)^\mu_\mu$ is scheme dependent~\cite{Hatta:2018sqd}. At one loop, the quark part simply accounts for the $n_f$ term in the beta function. } 
\beq
\Theta^\mu_\mu =\sum_f (\Theta_f)^\mu_\mu + (\Theta_g)^\mu_\mu= \frac{\beta(g)}{2g}F^2 = -\frac{\alpha_s}{8\pi}\left(\frac{11N_c}{3}-\frac{4T_Rn_f}{3}\right)F^2+{\cal O}(\alpha_s^2) \, .
\eeq
These observations suggest that the poles in~(\ref{offsym}) are physically there, at least in Compton scattering, and may carry profound implications possibly touching the issue of nonperturbative mass generation in QCD.

Let us now  explore the consequences of~(\ref{offsym}). 
Treating~(\ref{comp}) as a fully off-forward amplitude, we find from the operator product expansion
\beq
{\rm Im}T_{\rm sym}^{\mu\nu}
&=& \sum_f\frac{e_f^2}{4P^+}\left[\frac{4x_B^2}{Q^2}\left(P^\mu -\frac{P\cdot q}{q^2}q^\mu\right)\left(P^\nu-\frac{P\cdot q}{q^2}q^\nu\right) -g^{\mu\nu}+\frac{q^\mu q^\nu}{q^2}\right]\nonumber \\
&& \times\bar{u}(P_2)\left[\bigl(H^{\rm bare}_f(x_B,\xi,l^2)-H^{\rm bare}_{f}(-x_B,\xi,l^2)\bigr)\gamma^+ +\frac{i\sigma^{+\lambda}l_\lambda}{2M}\bigl(E^{\rm bare}_f(x_B,\xi,l^2)-E^{\rm bare}_{f}(-x_B,\xi,l^2)\bigr)\right]
u(p_1) \nonumber \\[0.2cm]
&& + {\cal O}(\alpha_s)+{\cal O}(1/Q^2) \, , \label{coll}
\eeq
where, as before, $H_f^{\rm bare}$ and $E_f^{\rm bare}$ are the `bare' versions of the standard unpolarized quark GPDs $H_f$ and $E_f$. In the forward limit, $H_f(x)=q_f(x)$ and  $-H_f(-x)=\bar{q}_f(x)$ are the quark and antiquark PDFs. According to QCD factorization theorems~\cite{Collins:1998be,Ji:1998xh}, higher-order corrections to~(\ref{coll}) not suppressed by $1/Q^2$ consist only of the twist-two quark and gluon GPDs to all orders. However, our result~(\ref{offsym}) shows that there are  diverging corrections already in the ${\cal O}(\alpha_s)$ terms which cannot be simply dismissed as higher twist, nor  be absorbed into the (usual) renormalization of twist-two GPDs. From the Gordon identity
\beq
\frac{\bar{u}(P_2)u(P_1)}{2M} =\frac{1}{2P^+}  \bar{u}(P_2)\left[\gamma^+-\frac{i\sigma^{+\lambda}l_\lambda}{2M}\right]u(P_1) \, ,
\eeq
we see that the pole affects both $H_f$ and $E_f$, with opposite signs. This raises the concern that factorization is violated also in the symmetric part of the Compton amplitude.

Nevertheless, the Compton amplitude must have a smooth forward limit.  Similar to the antisymmetric case discussed above (cf.~(\ref{canc})), we absorb the pole  in~(\ref{offsym}) into the definitions of the GPDs 
\beq
\sum_f e_f^2x_B(H_f(x_B,\xi,l^2)-H_{f}(-x_B,\xi,l^2)) &=& \sum_f e_f^2x_B(H^{\rm bare}_f(x_B,\xi,l^2)-H^{\rm bare}_{f}(-x_B,\xi,l^2)) \nn &&  +\frac{\alpha_s}{2\pi}\Bigg(\sum_f e_f^2\Bigg) 
\frac{x_B}{l^2}C^{\rm anom}\otimes' {\cal F}(x_B,\xi,l^2) \, ,
\eeq
\beq
\sum_f e_f^2x_B(E_f(x_B,\xi,l^2)-E_{f}(-x_B,\xi,l^2)) &=& \sum_f e_f^2x_B(E^{\rm bare}_f(x_B,\xi,l^2)-E^{\rm bare}_{f}(-x_B,\xi,l^2)) \nn &&  -\frac{\alpha_s}{2\pi}\Bigg(\sum_f e_f^2\Bigg) 
\frac{x_B}{l^2}C^{\rm anom}\otimes' {\cal F}(x_B,\xi,l^2) \, ,
\eeq
and let it cancel with nonperturbative, pre-existing poles in $H^{\rm bare}_f$ and $E^{\rm bare}_f$:
\begin{align}
\begin{split}
\sum_f e_f^2x_B(H^{\rm bare}_f(x_B,\xi,l^2)-H^{\rm bare}_{f}(-x_B,\xi,l^2)) & \approx -\frac{\alpha_s}{2\pi}\Bigg(\sum_f e_f^2\Bigg) 
\frac{x_B}{l^2}C^{\rm anom}\otimes' {\cal F}(x_B,\xi,l^2=0) \, ,\label{e:H_E_pole} \\
\sum_f e_f^2x_B(E^{\rm bare}_f(x_B,\xi,l^2)-E^{\rm bare}_{f}(-x_B,\xi,l^2)) & \approx \frac{\alpha_s}{2\pi}\Bigg(\sum_f e_f^2\Bigg)\frac{x_B}{l^2}C^{\rm anom}\otimes' {\cal F}(x_B,\xi, l^2=0) \, .
\end{split}
\end{align}
Again, this cancellation must occur for all values of $x_B$ and $Q^2$ in order to save the QCD factorization of the Compton amplitude. 
The second moments of the `renormalized' GPDs thus defined are  
\begin{align}
\begin{split}
\int_0^1 dx_B x_B  (H_f(x_B,\xi,l^2)-H_f(-x_B,\xi,l^2)) & = \int_{-1}^1 dx_B x_B H_f(x_B,\xi,l^2) = A_f(l^2) + \xi^2 D_f(l^2) \, ,\label{poly0} \\
\int_0^1 dx_B x_B (E_f(x_B,\xi,l^2)-E_f(-x_B,\xi,l^2)) & = \int_{-1}^1 dx_B x_B E_f(x,\xi,l^2) = B_f(l^2) - \xi^2 D_f(l^2) \, ,
\end{split}
\end{align}
where $A_f,B_f,D_f$ are the gravitational form factors defined by the off-forward matrix element of the (quark part of the) QCD energy momentum tensor~\cite{Ji:1996ek},
\beq
\langle P_2|\Theta^{\mu\nu}_f|P_1\rangle =
\frac{1}{M}\bar{u}(P_2)\left[P^\mu P^\nu A_f  + (A_f+B_f)\frac{P^{(\mu}i\sigma^{\nu)\rho}l_\rho}{2}+\frac{D_f}{4}(l^\mu l^\nu -g^{\mu\nu}l^2) + M^2\bar{C}_f g^{\mu\nu}\right]u(P_1) \, . 
\label{gff}
\eeq
On the other hand,  the second moment of  the pole term contains the following integral:
\beq
&&\int_0^1 dx_B x_B \left[\int_{x_B}^1 \frac{dx}{x}K\left(\frac{x_B}{x},\frac{\xi}{x}\right)-\frac{\theta(\xi-x_B)}{2}\int_{-1}^1\frac{dx}{x}L\left(\frac{x_B}{x},\frac{\xi}{x}\right)\right] {\cal F}(x,\xi,l^2) \nonumber \\ &&\ 
= \frac{T_R}{6}\int_0^1 dx \frac{x}{1-\frac{\xi^2}{x^2}}\left(1-\frac{\xi^4}{x^4}\right) {\cal F}(x,\xi,l^2) = \frac{T_R}{12}\int_{-1}^1 dx x\left(1+\frac{\xi^2}{x^2}\right) {\cal F}(x,\xi,l^2) \, . \label{enc}
\eeq
Note that if it were not for the second term, the result would contain infinitely many powers of $\xi$ from the expansion
\beq
\frac{1}{1-\frac{\xi^2}{x^2}} = 1+\frac{\xi^2}{x^2}+\frac{\xi^4}{x^4}+\ldots \, , \label{tranc}
\eeq
which would violate the polynomiality of GPD moments. In fact, the only role of the second term is to subtract all the higher-order terms $\xi^n$ with $n\ge 4$. The final result is thus quadratic in $\xi$, consistent  with the 
right-hand sides of~(\ref{poly0}). A cancellation of this sort is known in the GPD literature, see~\cite{White:2001pu}. Interestingly, our result~(\ref{whitec}) coincides with the `simple model' employed in~\cite{White:2001pu}.

The last integral in (\ref{enc}) reads 
\beq
\int_{-1}^1 dx x \left(1+\frac{\xi^2}{x^2}\right){\cal F}(x,\xi,l^2)= -\frac{4M}{(P^+)^2}\frac{\langle P_2|F^{\alpha\beta}(i\overleftrightarrow{D}^+)^2F_{\alpha\beta}|P_1\rangle}{\bar{u}(P_2)u(P_1)} -4\xi^2M  \frac{\langle P_2|F^{\mu\nu}F_{\mu\nu}|P_1\rangle}{\bar{u}(P_2)u(P_1)} \, ,
\eeq
where $\overleftrightarrow{D}^+\equiv\frac{\overrightarrow{D}^+ - \overleftarrow{D}^+}{2}$. 
We thus arrive at 
\begin{align}
\begin{split}
\sum_f  e_f^2 \bigl(A^{\rm bare}_f(l^2) +\xi^2D^{\rm bare}_f(l^2) \bigr) & \approx \frac{T_R\alpha_s}{12\pi l^2} \Bigg(\sum_f e_f^2\Bigg)   \left(\frac{\langle P|F^{\alpha\beta}(i\overleftrightarrow{D}^+)^2F_{\alpha\beta}|P\rangle}{(P^+)^2}+\xi^2 \langle P|F^2|P\rangle \right) \, , \label{AD} \\
\sum_f e_f^2 \bigl(B^{\rm bare}_f(l^2) -\xi^2 D^{\rm bare}_f(l^2) \bigr)  & \approx -\frac{T_R\alpha_s}{12\pi l^2} \Bigg(\sum_f e_f^2\Bigg) \left(\frac{\langle P|F^{\alpha\beta}(i\overleftrightarrow{D}^+)^2F_{\alpha\beta}|P\rangle}{(P^+)^2}  +\xi^2 \langle P|F^2|P\rangle \right) \, . 
\end{split}
\end{align}
Therefore, $A^{\rm bare}_f$, $B^{\rm bare}_f$ and $D^{\rm bare}_f$ all develop a pole $1/l^2$, but not in the linear combination $A_f^{\rm bare}+B_f^{\rm bare}=A_f+B_f$ relevant to the Ji sum rule~\cite{Ji:1996ek}. By construction, these poles cancel against the poles in~(\ref{offsym}) from the box diagrams.  The total form factors $A_f,B_f$ and $D_f$ are free of a massless pole, as they should.

We now see a clear connection between our results~(\ref{offsym}),~(\ref{AD}) and the `trace anomaly pole'~(\ref{af}) known in the literature after   taking into account the factor of 2 from~(\ref{def}) (and also the sign difference between (\ref{offsym}) and (\ref{AD})). 
This shows that the pole in the box diagram originates from the QCD trace anomaly. The ${\cal O}(\xi^2)$ terms match straightforwardly, but for  the ${\cal O}(\xi^0)$ terms there is an additional complication  due to the  convolution integral in $x$.

Let us finally speculate on the origin of the nonperturbative poles introduced in~(\ref{e:H_E_pole}) in an {\it ad hoc} way. 
Just like the massless pole in $g_P$ is induced by the would-be Nambu-Goldstone boson of unbroken $U_A(1)$ symmetry, the $\eta_0$ meson, those in the gravitational form factors might be induced by the would-be massless  tensor ($2^{++}$) and scalar ($0^{++}$) glueballs $G^{(2)}_0$ and $G^{(0)}_0$ in conformally symmetric QCD. Indeed, the exchange of a massless tensor glueball gives, schematically,  
\beq
\langle P_2|\Theta^{\mu\nu}|P_1\rangle \sim \langle P_2|P_1 G^{(2)}_0\rangle \frac{1}{l^2}\langle G^{(2)}_0|\Theta^{\mu\nu}|0\rangle &\sim& g^{(2)}_{GNN}f^{(2)}_{G}\frac{\bar{u}(P_2)P^{(\alpha}\gamma^{\beta)} u(P_1) }{l^2}\sum_{\rm pol} \epsilon_{\alpha\beta}\epsilon^{\mu\nu}\nn
&\sim& g^{(2)}_{GNN}f^{(2)}_{G} \frac{1}{l^2}\bar{u}(P_2)P^{(\mu}\gamma^{\nu)} u(P_1) \, ,
\eeq
where $g^{(2)}_{GNN}$ and $f^{(2)}_G$ are the effective glueball-nucleon coupling and the glueball decay constant, respectively. $\epsilon^{\mu\nu}$ is the glueball polarization tensor. Likewise, the exchange of a scalar glueball  gives\footnote{A careful analysis  shows that both $2^{++}$ and $0^{++}$ glueballs contribute to the $D$-form factor~\cite{Fujita:2022jus}, see also~\cite{Mamo:2019mka,Mamo:2021krl}. }
\beq
 \langle P_2|P_1 G^{(0)}_0\rangle \frac{1}{l^2}\langle G^{(0)}_0|\Theta^{\mu\nu}|0\rangle \sim g^{(0)}_{GNN}f^{(0)}_{G} \frac{ l^\mu l^\nu}{l^2}\bar{u}(P_2) u(P_1) \, .
\eeq
Comparing with (\ref{AD}), we find 
\beq
g^{(2)}_{GNN}f^{(2)}_G  \sim \frac{\langle P|\alpha_s F^{\alpha\beta}(i\overleftrightarrow{D}^+)^2F_{\alpha\beta}|P\rangle}{(P^+)^2} \, , \qquad     Mg^{(0)}_{GNN}f^{(0)}_G \sim \langle P|\alpha_s F^2|P\rangle \, .
\eeq
Of course, in  QCD conformal symmetry is explicitly broken by the trace anomaly and glueballs acquire nonzero masses.  After the cancellation, poles in the gravitational form factors are shifted to the physical glueball masses 
\beq
  \sum_f D_f(l^2) \approx M \frac{T_Rn_f \alpha_s}{6\pi l^2}\left(\frac{\langle P_2| F^2|P_1\rangle}{\bar{u}(P_2)u(P_1)}\Biggl|_{l^2=0} - \frac{\langle P_2| F^2|P_1\rangle}{\bar{u}(P_2)u(P_1)} \right)  \sim M\sum_n \frac{ g_{G_n NN}^{(0)}f_{G_n}^{(0)}}{l^2-m^2_{G_n}} \, , \label{dom}
\eeq
and similarly for the $A_f,B_f$ form factors. 
This is in accord  with the recently advocated `glueball dominance' picture~\cite{Fujita:2022jus} of the gravitational form factors (similar to the vector meson dominance of the electromagnetic form factors). 
Further investigations and discussions of these ideas are certainly needed.  

\section{Summary}
We have calculated the imaginary part of the one-loop quark box diagrams for the antisymmetric and symmetric parts of the Compton amplitude relevant to DVCS. The results in the forward limit correspond to the DIS structure functions. Our calculations are distinct from all previous work in the GPD literature in the sense that we have kept $t=l^{2}$ finite. This is mandatory in order to correctly isolate the infrared $1/l^{2}$ poles related to the chiral anomaly (in the antisymmetric channel) and the trace anomaly (in the symmetric channel). The need to keep $t=l^{2}$ finite was  pointed out in Refs.~\cite{Tarasov:2020cwl,Tarasov:2021yll} in the context of polarized DIS. Below we summarize our main findings:
\begin{itemize}
\item[---] \textit{Antisymmetric case}: Our main result is Eq.~(\ref{off}) which features (i) a pole $1/l^2$ coupled to the gluon fields through the pseudoscalar operator $F^{\mu\nu}\tilde{F}_{\mu\nu}$, (ii) the usual logarithmic term and the coefficient function in polarized DIS. We interpret this result in terms of the GPDs $\tilde{H}$ and $\tilde{E}$ by mapping~(\ref{off}) to the QCD  factorization formula for the Compton amplitude. The appearance of a genuine pole  $1/l^2$ is a serious concern in this context since such a term does not fit in this framework. However, in order to still justify factorization and to ensure a well-defined forward limit for the Compton amplitude, we have introduced a renormalized definition of the  GPD $\tilde{E}$~(\ref{canc}) in which this perturbative pole cancels with a nonperturbative pole present {\it a priori} in the `bare' GPD $\tilde{E}^{\rm bare}$. We have further postulated that it is the integral of this renormalized  $\tilde{E}$ that defines the pseudoscalar form factor $g_P$. This requirement has nevertheless led to the relation~(\ref{cf2}) which we view as correct and is consistent with the pole cancellation argument in~\cite{Jaffe:1989jz,Tarasov:2021yll}. 
\item[---] \textit{Symmetric case}:  Our main result is Eq.~(\ref{offsym}) which features (i) a pole $1/l^{2}$ coupled to the gluon fields through the scalar operator $F^{\mu\nu}F_{\mu\nu}$, (ii) the usual logarithmic term and the coefficient functions in unpolarized DIS. We again interpret this result in terms of the GPDs $H$ and $E$ by mapping~(\ref{offsym}) to the QCD factorization formula. To the best of our knowledge, the observation of such a genuine pole in the symmetric sector is new, but once again this raises  concerns for the validity of QCD factorization. Following the same line of reasoning as in the antisymmetric case, we have introduced  renormalized definitions of the GPDs $H$ and $E$ which undergo cancellation of the perturbative poles with the nonperturbative, pre-existing poles from massless glueball exchanges. The same cancellation occurs in the gravitational form factors, shifting massless poles to massive glueball poles (\ref{dom}). 
\end{itemize}

We emphasize that, while the above arguments are to some extent speculative, without such arguments the QCD factorization of the Compton amplitude is violated or at least requires revisions  in both the symmetric and antisymmetric parts. On the other hand, from these one-loop calculations  we do not foresee problems with the QCD factorization in inclusive DIS. In the polarized case, the pole term~(\ref{off}) has a distinct Lorentz structure $\propto l^\alpha$ from the $\Delta G$ term $\propto S^\alpha$. The cancellation of poles occurs entirely within the GPD $\tilde{E}$ sector while $g_1$ is renormalized in the usual way. In the unpolarized case, the pole vanishes after averaging over the gluon polarizations~(\ref{nopole}) appropriate to unpolarized DIS. (See however a caveat above~(\ref{47}).)

In the future, we plan to extend our calculation to the real part of the Compton amplitude as well as the quark-initiated diagrams in DVCS and DIS. The implications of our result for the gravitational form factors will also be  further investigated. We hope that our work will motivate further studies
that shed additional light on the role of anomalous contributions to DVCS and their potential impact on factorization.

\begin{acknowledgements}
We thank Andrey Tarasov and Raju Venugopalan for all the stimulating discussions that triggered this work and continued throughout the course of the work. We also thank Swagato Mukherjee and Jianwei Qiu for valuable discussions. S.~B. thanks Hiren Patel for helpful discussions. S.~B. and Y.~H. were supported by the U.S. Department of Energy under Contract No. DE-SC0012704, and also by  Laboratory Directed Research and Development (LDRD) funds from Brookhaven Science Associates. S.~B. has also been supported by the U.S. Department of Energy, Office of Science, Office of Nuclear Physics and Office of Advanced Scientific Computing Research within the framework of Scientific Discovery through Advance Computing (SciDAC) award Computing the Properties of Matter with Leadership Computing Resources. W.~V. has been supported by Deutsche Forschungsgemeinschaft (DFG) through the Research Unit FOR 2926 (project 430915220). He is also grateful to Brookhaven National Laboratory for hospitality during completion of part of this work.
\end{acknowledgements}

\appendix

\section{Analytical derivation of the pole term}
In the body of the paper, we have used the Mathematica package  `Package-X'~\cite{Patel:2015tea} to compute the imaginary part of the relevant Feynman diagrams. It is instructive to  demonstrate analytically how the pole term $1/l^2$ arises. This has been done in~\cite{Tarasov:2020cwl} in the `worldline' formalism. In this appendix, we rederive the same result from an ordinary Feynman diagram calculation.

For definiteness, we consider the antisymmetric case. 
Our approach is similar to that in~\cite{Carlitz:1988ab} except that we regularize the infrared singularity by off-forward momentum transfer $l^\mu$ instead of the current quark mass $m_q$. The relevant integral is  
\beq
  \epsilon^{\alpha\beta}_{\ \ \, \mu\nu} p_\beta\int \frac{d^4k}{(2\pi)^4}  \delta((q+k)^2)((k-p)^2) \frac{{\rm Tr}[\gamma^\mu (\Slash k+\Slash q)\gamma^\nu (\Slash k -\Slash l/2)\gamma^\rho (\Slash k-\Slash p)\gamma^\lambda (\Slash k+\Slash l/2)]}{(k-l/2)^2(k+l/2)^2} \epsilon_{1\rho}\epsilon^*_{2\lambda} \, , \label{a11}
\eeq
where we have already taken the imaginary part by putting the final state partons on shell. 
The delta function constraints can be solved as 
\beq
\delta((p-k)^2)\delta((q+k)^2)\approx \frac{\hat{x}}{2(1-\hat{x})Q^2} \Bigl(\delta(k^+-k_1^+)\delta(k^--k_1^-)+  \delta(k^+-k_2^+)\delta(k^--k_2^-) \Bigr) \, , \label{solve}
\eeq
where
\beq
&&k^+_{1,2} = \frac{p^+}{2}\left(1+\hat{x} \mp(1-\hat{x}) \sqrt{1-\frac{16\hat{x}k_\perp^2}{(1-\hat{x})(4Q^2-\hat{x}l^2)}}\right) \, ,\nn
&& k^-_{1,2} = -\frac{1}{16\hat{x}p^+}\left( 4Q^2+\hat{x}l^2\mp (4Q^2-\hat{x}l^2)  \sqrt{1-\frac{16\hat{x}k_\perp^2}{(1-\hat{x})(4Q^2-\hat{x}l^2)}}\right) \, .  \label{root}
\eeq
The pole comes from the first set of roots, $k_1^\pm$, in the $k_\perp \to 0$ region. In this region, we can approximate  
\beq
k_1^+\approx 
\hat{x}p^+ \, , \qquad 
k_1^-\approx \frac{-k_\perp^2}{2(1-\hat{x})p^+}-\frac{l^2}{8p^+} \, . \label{same}
\eeq
After performing the $k^\pm$ integrals using the delta functions, we are left with an integral of the form 
\beq
\int d^2k_\perp \frac{f(k_\perp)}{(k-l/2)^2(k+l/2)^2} = \int_0^1 da \int d^2k_\perp\frac{f(k_\perp)}{\left(k^2+\frac{l^2}{4} + (1-2a) k\cdot l\right)^2} \, . \label{left}
\eeq
Using 
\beq
&&k^2+\frac{l^2}{4} \approx \frac{-k_\perp^2}{1-\hat{x}}+\frac{(1-\hat{x}){l^2}}{4} \, , \nn
&& k^+l^- + k^-l^+ \approx 
\frac{(1-\hat{x})l^2\hat{\xi}}{4} + \frac{k_\perp^2 \hat{\xi}}{1-\hat{x}} \, , \label{xikeep} 
\eeq
we can evaluate (\ref{left}) as
\beq
&&\int_0^1 da \int^{\frac{1-\hat{x}}{4\hat{x}}Q^2} d^2k_\perp \frac{f(k_\perp)}{\left(\frac{k_\perp^2}{1-\hat{x}}-\frac{(1-\hat{x})l^2}{4} +(1-2a)\left(k_\perp \cdot l_\perp -\frac{(1-\hat{x})l^2\hat{\xi}}{4} -\frac{k_\perp^2 \hat{\xi}}{1-\hat{x}} \right) \right)^2} \nn
&&= \int_0^1 da \int^{\frac{1-\hat{x}}{4\hat{x}}Q^2}d^2k'_\perp \frac{f\left(k'_\perp- \frac{(1-\hat{x})(1-2a)}{2(1-\hat{\xi}(1-2a))}l_\perp\right)}{\left(\frac{1-\hat{\xi}(1-2a)}{1-\hat{x}}k'^2_\perp - \frac{(1-\hat{x})a(1-a)}{1-\hat{\xi}(1-2a)}l^2\right)^2} \nn
&& \approx \pi \int_0^1 da  \frac{f\left(-\frac{1}{2} (1-\hat{x})(1-2a)l_\perp\right)}{-a(1-a)l^2} \, , \label{aint}
\eeq
where we have neglected $k'_\perp$ in the numerator since the pole comes from the infrared. We have also  used the relation
\beq
l^2= -\frac{\vec{l}^{\;2}_\perp}{1-\hat{\xi}^2}\, , 
\label{e:spacelike}
\eeq
 to obtain the second line.  
This relation is crucial to restore Lorentz covariance which was not manifest after performing the $k^\pm$-integrations. 
The apparent singularities in~(\ref{aint}) at $a=0,1$ are innocuous because the trace in the numerator gives  
\beq
f(...) \sim  a(1-a) p\cdot q(1-\hat{x})^2 (\epsilon_1\cdot l \epsilon^{\alpha\epsilon^*_2 l p} -\epsilon^*_2\cdot l \epsilon^{\alpha \epsilon_1 l p}) +\ldots \, .  \label{tensor}
\eeq
This term is quadratic in $l$, so it can be easily  missed if one is not careful in taking the forward limit.
With the help of the Schouten identity, the tensors in~(\ref{tensor}) can be rewritten as 
\beq
\epsilon_1 \cdot l \epsilon^{\alpha \epsilon^*_2 lp}-\epsilon_2^*\cdot l e^{\alpha\epsilon_1 l p} +l^2\epsilon^{\alpha \epsilon_1 \epsilon^*_2 p} = l^\alpha \epsilon^{\epsilon_1 \epsilon^*_2lp} \, ,
\eeq
where we included a non-pole term proportional to $l^2$. This term is beyond the accuracy of the present schematic derivation, but it must be there due to gauge invariance.  
Combining with the prefactor $\frac{\hat{x}}{(1-\hat{x})Q^2}$ from~(\ref{solve}), we recover the anomaly pole
\beq
\frac{p\cdot q}{Q^2}\hat{x}(1-\hat{x})\frac{l^\alpha  }{l^2}\epsilon^{\epsilon_1 \epsilon^*_2 lp} \sim (1-\hat{x})\frac{l^\alpha }{l^2}\epsilon^{\epsilon_1 \epsilon^*_2 lp} \, .
\eeq

It is easy to see that  the same pole arises in the polarized quark GPD of a single gluon
\beq
{\cal A}_g^+\equiv p^+\int \frac{dz^-}{4\pi}e^{i\hat{x}p^+z^-}\langle p_2|\bar{\psi}(-z^-/2)\gamma^+\gamma_5\psi(z^-/2)|p_1\rangle \, ,
\eeq
evaluated to one loop. For $0<\hat{x}<1$, 
this can be obtained by replacing 
$\epsilon^{+\beta}_{\ \ \ \mu\nu}p_\beta \gamma^\mu (\Slash k+\Slash q)\gamma^\nu \to \gamma^+\gamma_5$ and $\delta((q+k)^2)\to \delta(k^+-\hat{x}p^+)$ in~(\ref{a11}) so that the roots~(\ref{same}) remain unchanged. 
The calculation is simpler and here we report on the complete analytic result in the limit $l^\mu\to 0$ 
\beq
{\cal A}_g^+ \approx  \frac{\alpha_s}{2\pi}T_R \left[(2\hat{x}-1)\left(\ln  \frac{4\Lambda^2}{-l^2(1-\hat{x})^2}-1\right)\otimes \delta(1-\hat{x})i\epsilon^{+p\epsilon_2^*\epsilon_1}+ \frac{2il^+}{l^2}(1-\hat{x})\otimes \delta(1-\hat{x})\epsilon^{\epsilon_1 \epsilon_2^*lp}\right] \, . \label{polepole}
\eeq
The cutoff $\Lambda$ of the transverse momentum integral has to be chosen such that the $\hat{x}$-integral of the first term vanishes. This is analogous to fixing the  ambiguity of the triangle diagram for $\langle J_5^\mu\rangle$ by requiring gauge invariance (vector current conservation). In particular, with the choice $\Lambda^2=\frac{1-\hat{x}}{4\hat{x}}Q^2$ (cf.,~(\ref{aint})), 
Eq.~(\ref{polepole}) has exactly the same structure as~(\ref{off}) including even the finite terms.  
Adding the contribution from the region $-1<\hat{x}<0$ and integrating over $\hat{x}$, one then recovers~(\ref{trian}).

Let us finally comment on the connection to semi-inclusive DIS (SIDIS). Since the inclusive cross section has a pole, there must be a trace of it in the SIDIS cross section.  This can be seen by  undoing the integral over $k_\perp\approx k'_\perp$ in~(\ref{aint}) which is the transverse component of the  final-state quark  momentum $k_q=q+k$.  $k_\perp$ is related to the standard variable $\hat{z}=\frac{p\cdot k_q}{p\cdot q}$ in SIDIS as  
\beq
\frac{k^2_\perp}{1-\hat{x}} = \hat{z}(1-\hat{z}) \frac{Q^2}{\hat{x}}
\eeq
so that $dk^2_\perp \sim |1-2\hat{z}|d\hat{z}$. We thus see that the pole comes from the integral
\beq
\frac{\pi}{2} \frac{\hat{x}(1-\hat{x})}{Q^2}\int_0^{1} d\hat{z} \frac{|1-2\hat{z}|}{(\hat{z}(1-\hat{z}) - \hat{x}(1-\hat{x})a(1-a) l^2/Q^2)^2}\approx \frac{\pi}{-a(1-a)l^2} \, .
\eeq
This shows that the two end points $\hat{z}\to 0,1$ equally contribute to the pole. We again emphasize that to access this pole in SIDIS, one has to keep terms quadratic in $l^\mu$ in the numerator.

\bibliography{references}

\end{document}